\documentclass[aps,prl,reprint,superscriptaddress,longbibliography]{revtex4-1}
\usepackage{graphicx}
\usepackage{rotating}
\usepackage{color}
\usepackage{soul}
\usepackage{caption}
\usepackage{subcaption}
\usepackage{amsmath,amssymb}

\begin{document}

\title{Microscopic Engine Powered by Critical Demixing}

\author{Falko Schmidt}
\affiliation{Department of Physics, University of Gothenburg, 41296 Gothenburg, Sweden}
\affiliation{Soft Matter Lab, Department of Physics and UNAM -- National Nanotechnology Research Center, Bilkent University, Ankara 06800, Turkey}
\affiliation{Peter-Debye-Institute for Soft Matter Physics, Faculty of Physics and Earth Science, University of Leipzig, 04103 Leipzig, Germany}

\author{Alessandro Magazz\`u}
\affiliation{Department of Physics, University of Gothenburg, 41296 Gothenburg, Sweden}
\affiliation{Soft Matter Lab, Department of Physics and UNAM -- National Nanotechnology Research Center, Bilkent University, Ankara 06800, Turkey}

\author{Agnese Callegari}
\affiliation{Soft Matter Lab, Department of Physics and UNAM -- National Nanotechnology Research Center, Bilkent University, Ankara 06800, Turkey}

\author{Luca Biancofiore}
\affiliation{Department of Mechanical Engineering, Bilkent University, Ankara 06800, Turkey}

\author{Frank Cichos}
\affiliation{Peter-Debye-Institute for Soft Matter Physics, Faculty of Physics and Earth Science, University of Leipzig, 04103 Leipzig, Germany}

\author{Giovanni Volpe}
\affiliation{Department of Physics, University of Gothenburg, 41296 Gothenburg, Sweden}
\affiliation{Soft Matter Lab, Department of Physics and UNAM -- National Nanotechnology Research Center, Bilkent University, Ankara 06800, Turkey}

\date{\today}


\begin{abstract}
By converting energy into mechanical work, engines play a central role in most biological and technological processes. 
In particular, within the current trend towards the development of nanoscience and nanotechnology, microscopic engines have been attracting an ever-increasing interest.
On the one hand, there has been a quest to understand how biological molecular motors work. 
On the other hand, several approaches have been proposed to realize artificial microscopic engines, which have been powered by the transfer of light momentum, by external magnetic fields, by \emph{in situ} chemical reactions, or by the energy flow between hot and cold heat reservoirs, in scaled-down versions of macroscopic heat engines.
Here, we experimentally demonstrate a microscopic engine powered by the local reversible demixing of a critical mixture.
We show that, when an absorbing microsphere is optically trapped by a focused laser beam in a sub-critical mixture, it is set into rotation around the optical axis of the beam because of the emergence of diffusiophoretic propulsion; this behavior can be controlled by adjusting the optical power, the temperature, and the criticality of the mixture.
Given its simplicity, this microscopic engine provides a powerful tool to power micro- and nanodevices. Furthermore, since many biological systems are tuned near criticality,
this mechanism might already be at work within living organisms, for example in proteins and in cellular membranes.

\end{abstract}

\maketitle

\section{Introduction}

Engines hold the central stage in many natural and technological systems as devices capable of converting energy into mechanical work. During the last few decades, a lot of effort has gone into miniaturizing engines down to nanoscopic length scales for applications in nanoscience and nanotechnology \cite{Browne2006,Abdelmohsen2014}. Differently from their macroscopic counterparts, microscopic engines are not completely deterministic, because they operate on energy scales where thermal fluctuations become relevant, and, therefore, need to be treated within the context of stochastic thermodynamics \cite{seifert2012,martinez2017colloidal}.

Several approaches have been proposed to realize microsopic engines capable of  performing work. For example, microrotators have been realized by transferring the orbital and spin momentum of light to microscopic particles \cite{Simpson97,Friese1998} or by employing rotating magnetic fields \cite{Biswal2004,ghosh2009controlled,eickenberg2013}. Several prototypes of microscopic heat engines have been realized exploiting the nucleation of a vapor bubble inside a silicon microcavity, some of them down to a working volume of only $0.6\,{\rm mm^3}$ \cite{Kao2007,Lee2011,Percy2014}.  More recently, optically trapped particles have been employed to reproduce microscopic versions of the Stirling and Carnot cycles, and to study their stochastic thermodynamic properties \cite{Blickle2012, Martinez2016}. Also, a microscopic steam engine has been developed based on the periodic generation of cavitation bubbles by an optically trapped particle \cite{QuintoSu2014}. 

Here, we introduce a new mechanism to power a microscopic engine which relies on the local and reversible demixing of a critical mixture surrounding a microparticle. In particular, we show that an absorbing spherical microparticle, dissolved in a critical binary mixture and optically trapped, is able to perform rotational motion around the beam waist and to produce work thanks to the local demixing generated by the (slight) increase of the temperature of the solution when the particle approaches the focal point. The properties of this microscopic engine can be controlled by adjusting the optical power, the temperature, and the criticality of the binary mixture. Differently from the artificial microscopic engines mentioned above, the microscopic engine we propose here does not rely on the transfer of (angular) momentum from an external source (e.g. from circularly polarized light fields \cite{Simpson97,Friese1998}, high-order laser beams \cite{He1995}, or magnetic fields \cite{ghosh2009controlled}) or on a flow of energy from a hot reservoir to a cold reservoir (e.g. microscopic Stirling \cite{Blickle2012}, Carnot \cite{Martinez2016}, and steam engines \cite{QuintoSu2014}).

\section{Results}

\begin{figure*}
\includegraphics[width=\textwidth]{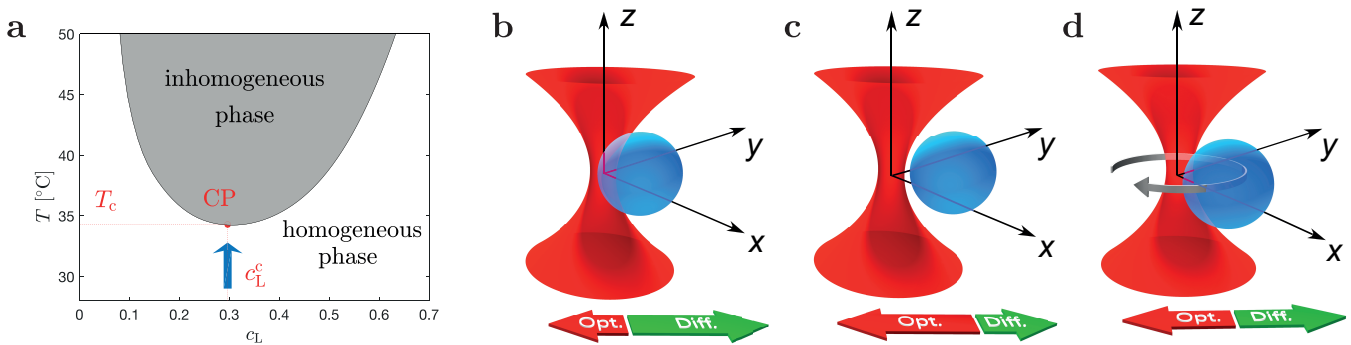}
\caption{
\textbf{Critical engine working principle.}
(a) Phase diagram of the water--2,6-lutidine mixture featuring a lower critical point (CP) at the bottom of the coexistence line (solid line). The system is prepared at the critical lutidine mass fraction $c_{\rm c} = 0.286$ and at a temperature $T_0$ significantly (several degrees) below the critical temperature $T_{\rm c} \approx 34^{\circ}{\rm C}$ (arrow).
(b-d) An optically trapped microsphere experiences a harmonic restoring optical force that attracts it towards the center of the optical trap near the focal spot (red arrows). If the particle absorbs the illumination light, the temperature of the surrounding fluid increases generating an asymmetric temperature profile, which is hotter on the side of the particle closer to the focal spot. Because of the criticality of the mixture, the temperature gradient, in turn, generates a concentration gradient surrounding the particle and, eventually, a diffusiophoretic drift (green arrows). (b) When the particle is close to the focal spot, the diffusiophoretic drift dominates. (c) Instead, when the particle is far from the focal spot, the optical-force-induced drift dominates. (d) The equilibrium position along the radial direction lies  where the optical and diffusiophoretic drifts balance each other and depends on the value of $T_0$ and on the light intensity; here the presence of small asymmetries in the temperature and demixing profile around the particle make the particle rotate around the optical axis (gray arrow).
}
\label{fig1}
\end{figure*}

We consider an absorbing microscopic sphere (silica with iron oxide inclusions, radius $R = 1.24\pm0.04 \,{\rm \mu m}$) held by an optical tweezers in a binary mixture of water and 2,6-lutidine at the critical lutidine mass fraction $c_{\rm c} = 0.286$ with a lower critical point at the temperature $T_{\rm c} \approx 34^{\circ}{\rm C}$ (see the phase diagram in Fig.~\ref{fig1}a) \cite{Grattoni1993}. When its temperature $T$ is below $T_{\rm c}$, the mixture is homogeneous and behaves like a normal fluid. As $T$ approaches $T_{\rm c}$ (arrow in Fig.~\ref{fig1}a), density fluctuations emerge. Finally, as $T$ exceeds $T_{\rm c}$, the mixture demixes into water-rich and lutidine-rich phases. 

The setup is based on an optical tweezers with a near-infrared laser beam ($\lambda = 976\,{\rm nm}$) built on a homemade inverted microscope; the sample stage temperature ($T_0$) is measured and stabilized with a feedback controller to within $\pm 3 \,{\rm mK}$ (see supplementary Fig.~S1 and the ``Setup" section in the methods). The particle's position is tracked by digital video microscopy at $296\,{\rm fps}$. Rhodamine B is added to the solution to indicate the phase separation of water and lutidine: As Rhodamine B is water soluble and fluoresces around $600\,{\rm nm}$, the detected intensity provides information about the relative water content.

The microsphere is attracted by optical forces (red arrows in Figs.~\ref{fig1}b-d and supplementary Figs.~S2a-c) towards the center of the optical trap, which is near the focal spot. Due to the presence of iron oxide inclusions, the silica microsphere absorbs part of the light of the trapping beam and converts it into heat, producing a local increase of the temperature (supplementary Figs.~S2d-f). Since the side of the particle closer to the focal spot is hotter, an asymmetric temperature profile arises in the liquid surrounding the particle. Because of the criticality of the mixture, this temperature profile induces a concentration profile around the particle (supplementary Figs.~S2g-i). Finally, the presence of this concentration profile generates a diffusiophoretic motion \cite{Wurger2015} of the particle away from the focal spot (green arrows in Figs.~\ref{fig1}b-d and supplementary Figs.~S2a-c). Since the heating produced by the particle depends on the light intensity, it decreases as the particle moves radially away from the focal spot and, therefore, also the associated chemical gradient and diffusiophoretic motion decrease. As a consequence, the particle settles off-axis, where the drift due to optical forces is balanced by the diffusiophoretic drift, i.e. where the total radial force acting on the particle is zero (Fig.~\ref{fig1}d and supplementary Fig.~S2b). Importantly, there are small asymmetries in the composition of the particle that induce asymmetries in the temperature and demixing profiles around the particle and, consequently, make the particle rotate around the optical axis (gray arrow in Fig.~\ref{fig1}d and supplementary Fig.~S2b). Several parameters play a role in the workings of this engine, in particular the power $P$ of the trapping beam and ambient temperature $T_0$ of the surrounding fluid. 

\subsection{Critical engine operation as a function of power}

\begin{figure}[h]
\includegraphics[width=0.5\textwidth]{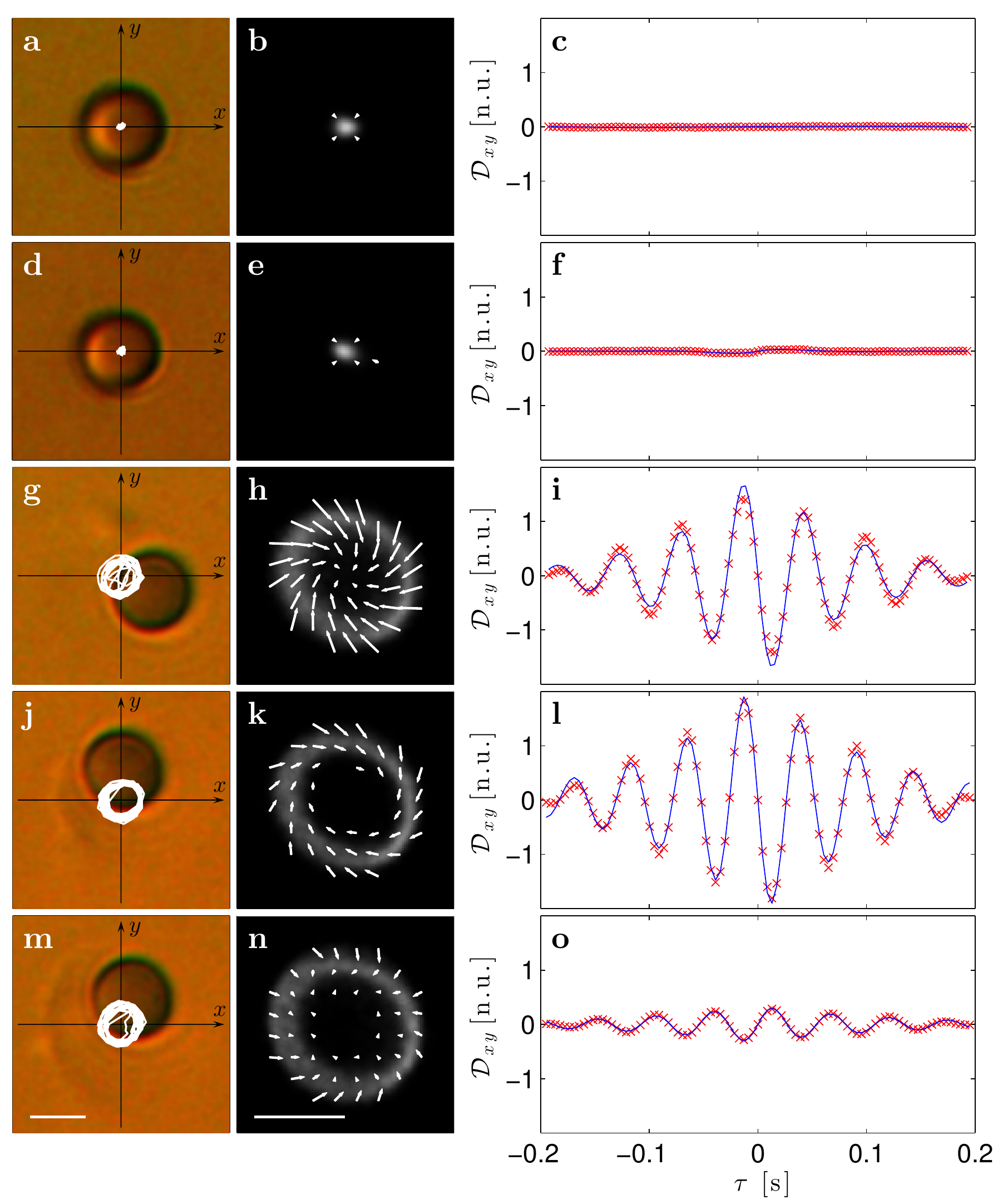}
\caption{
\textbf{Engine performance as a function of laser power.} 
The ambient temperature of the sample is fixed at $T_0 = 26^{\circ}{\rm C}$, while the laser power at the optical trap is (a-c) $P = 0.6$, (d-f) $1.5$, (g-i) $2.1$, (j-l) $2.7$, and (m-o) $3.2\,{\rm mW}$. (a,d,g,j,m) Bright-field images of the particle with $0.6\,{\rm s}$ trajectories represented by white solid lines. (b,e,h,k,n) Velocity drift fields (white arrows) and particle position probability distributions (background color, brighter colors represent higher probability distribution). (c,f,i,l,o) Experimental (red symbols) and fitted (blue solid lines) differential cross-correlation functions $\mathcal{D}_{xy}$ in the $xy$-plane. The white bars in (m) and (n) correspond to $1\,{\rm \mu m}$.
}
\label{fig2}
\end{figure}

We start by setting the ambient temperature of the sample to $T_0 = 26^{\circ}{\rm C}$. At low laser power ($P = 0.6\,{\rm mW}$ at the trap), the particle lingers around the center of the optical trap (solid line in Fig.~\ref{fig2}a). From the trajectory we calculate the velocity drift field, which indicates the direction and magnitude of the particle's velocity as a function of its position, and the differential cross-correlation function $\mathcal{D}_{xy}$, whose value indicates the magnitude of the rotational component of the force fields \cite{Volpe2007} (see ``Data analysis" section in the methods). The resulting drift field (white arrows in Fig.~\ref{fig2}b) points towards the center of the optical trap and $\mathcal{D}_{xy}$ (experimental symbols and theoretical fitting in Fig.~\ref{fig2}b) is negligible, showing that there is no cross-correlation between the movement along the $x$- and $y$-directions, which means that non-conservative forces and therefore rotation are absent. All these results are consistent with the behavior of an optically trapped particle in a non-critical medium \cite{jones2015optical}; this is expected because at low power the heating is not enough to approach $T_{\rm c}$ and, thus, to produce a demixing profile around the particle. 

Even increasing the power up to $P = 1.5\,{\rm mW}$, the particle is still constrained around the center of the optical trap (Fig.~\ref{fig2}d), the drift field points towards the center of the trap (Fig.~\ref{fig2}e), and non-conservative forces are negligible (Fig.~\ref{fig2}f). This is because the increase of temperature due to the light absorption is not sufficient to reach $T_{\rm c}$.

Increasing the optical power to $P = 2.1\,{\rm mW}$, the temperature of the solution surrounding the particle reaches $T_{\rm c}$ leading to a local demixing of the critical mixture (Fig.~\ref{fig2}g). The brighter regions surrounding the particle in Fig.~\ref{fig2}g correspond to water-rich regions, where the fluorescence of the Rhodamine B dye added to the solution is enhanced. The resulting concentration gradient induces a diffusiophoretic drift that pushes the particle radially outwards from the center, where it reaches a radial equilibrium position and starts rotating inside a toroidal region around the optical axis (see Figs.~\ref{fig1}b-d and associated discussion). This rotational motion can be seen from the particle trajectory (solid line in Fig.~\ref{fig2}g) and, more quantitatively, from the drift field (white arrows in Fig.~\ref{fig2}h), the particle probability distribution (background shading in Fig.~\ref{fig2}h), and $\mathcal{D}_{xy}$ (Fig.~\ref{fig2}i), which shows the sinusoidal behavior characteristic of the presence of rotational force fields. We remark that the rotational motion of the particle is due to structural asymmetries in the particle itself (e.g. in the distribution of the iron oxide and in its surface properties) and in the ensuing temperature gradient profile. Sometimes it can be observed that the rotation stops, the particle moves towards the center of the trap, and subsequently starts rotating again in the opposite direction.

Increasing the power even further to $P = 2.7\,{\rm mW}$, the particle performs constant revolutions around the optical trap center and its radial equilibrium position reaches about $1\,{\rm \mu m}$ (Fig.~\ref{fig2}j). The resulting drift field shows clearly a steady clockwise rotation (Fig.~\ref{fig2}k) and $\mathcal{D}_{xy}$ increases its amplitude (Fig.~\ref{fig2}l).  We observe that the particle performs continuous rotations around the optical axis without changing its direction during the measurement.

As the power is further increased to $P = 3.2\,{\rm mW}$, the particle not only rotates around but passes occasionally through the optical trap center and changes randomly its direction of rotation (Fig.~\ref{fig2}m).  The higher laser power makes the temperature of the critical mixture surrounding the particle significantly exceed $T_{\rm c}$, leading to a disruption of the balance between optical and diffusiophoretic drifts. As a consequence, the particle is attracted towards the optical trap center before being pushed radially outwards by diffusiophoresis and returning to rotational motion.  This leads to a partial disruption of the rotational component of the drift field (Fig.~\ref{fig2}n) and to a decrease of $\mathcal{D}_{xy}$ (Fig.~\ref{fig2}o).

When the power is further raised, the diffusiophoretic motion exceeds the optical trapping potential and therefore the particle escapes from the optical trap.

\subsection{Critical engine operation as a function of ambient temperature}

\begin{figure}[h!]
\includegraphics[width=0.5\textwidth]{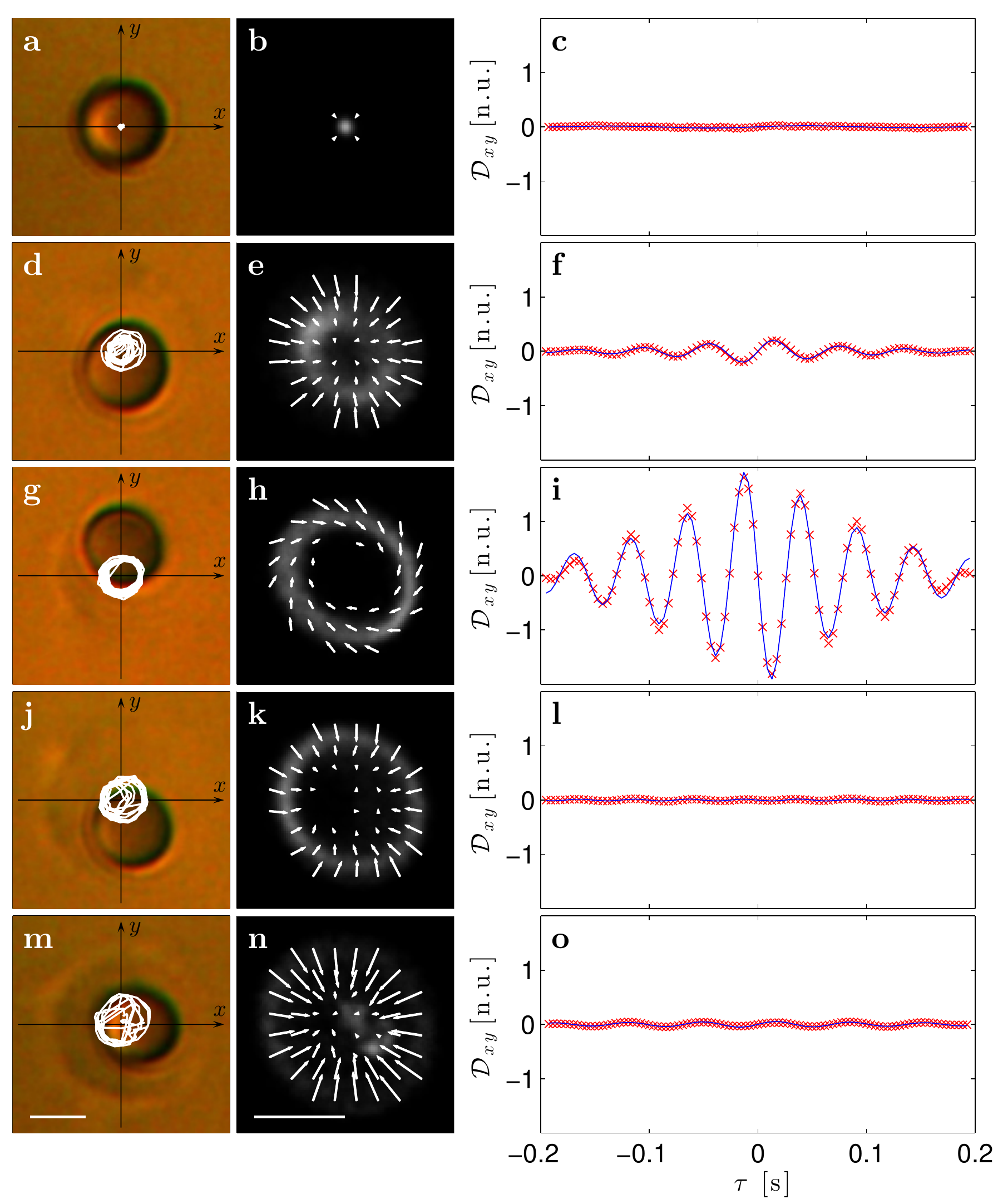}
\caption{
\textbf{Engine performance as a function of ambient temperature.}
The laser power at the optical trap is fixed at $P = 2.7\,{\rm mW}$, while the ambient temperature of the sample is (a-c) $T_0 = 20$, (d-f) $24$, (g-i) $26$, (j-l) $27$, and (m-o) $28^{\circ}{\rm C}$. (a,d,g,j,m) Bright-field images of the particle with $0.6\,{\rm s}$ trajectories represented by white solid lines. (b,e,h,k,n) Velocity drift fields (white arrows) and particle position probability distributions (background color, brighter colors represent higher probability distribution). (c,f,i,l,o) Experimental (red symbols) and fitted (blue solid lines) differential cross-correlation functions $\mathcal{D}_{xy}$ in the $xy$-plane. The white bars in (m) and (n) correspond to $1\,{\rm \mu m}$. (g-i) are the same data as Fig.~\ref{fig2}j-l.
}
\label{fig3}
\end{figure}

Now we fix the laser power at $P = 2.7\,{\rm mW}$ and study the behavior of the critical engine as a function of the ambient temperature $T_{0}$. 

At $T_{0}=20^{\circ}{\rm C}$, the system behaves as a standard optical trap: the particle trajectory lingers near the center of the optical trap (solid line in Fig.~\ref{fig3}a); the drift field points towards the optical trap center (white arrows in Fig.~\ref{fig3}b); and $\mathcal{D}_{xy}$ is negligible (Fig.~\ref{fig3}c).

As $T_0$ is increased to $24^{\circ}{\rm C}$, the particle explores a larger area around the center of the optical trap (Fig.~\ref{fig3}d) and rotational forces start emerging as shown by the drift field (Fig.~\ref{fig3}e) and by $\mathcal{D}_{xy}$ (Fig.~\ref{fig3}f). The diffusiophoretic drift generated by the particle is however not large enough to set the particle into constant rotation around the optical axis, as can be seen from the fact that the particle trajectory occasionally crosses the center of the trap (solid line in Fig.~\ref{fig3}d).

At $T_0 = 26^{\circ}{\rm C}$, the particle steadily rotates around the optical trap center (Fig.~\ref{fig3}g), leading to a well-defined rotational drift field (Fig.~\ref{fig3}h) and to an increase of $\mathcal{D}_{xy}$ (Fig.~\ref{fig3}i). 

At even higher temperatures (i.e. $T_0 = 27^{\circ}{\rm C}$ and $T_0 = 28^{\circ}{\rm C}$), the particle starts to occasionally change its orientation passing through the optical trap center before returning to its revolutionary motion (Figs.~\ref{fig3}j,m). With increasing $T_0$, the drift field looses the rotational component (Figs.~\ref{fig3}k,n) and $\mathcal{D}_{xy}$ decreases (Figs.~\ref{fig3}l,o), which are clear signatures of decreasing rotational forces.

A further increase in temperature leads to the particle being pushed away from the optical trap by the presence of overwhelming diffusiophoretic drifts.

\subsection{Critical engine operation as a function of the criticality of the mixture}

Besides $P$ and $T_0$, also the criticality of the binary mixture affects the behavior of the critical engine. All results presented so far were obtained with a binary mixture adjusted at the critical concentration corresponding to the critical lutidine mass fraction $c_{\rm c} = 0.286$. In order to explore the dependence on the criticality of the mixture, we repeated the measurements at an off-critical lutidine mass fraction of $0.236$ as a function of both $P$ and $T_0$. We did not observe any rotational behavior even at high power ($P>6\,{\rm mW}$) and we therefore concluded that the criticality of the mixture is an essential ingredient for the engine to work. This can be explained from the fact that when the mixture is off-critical, water and lutidine undergo a more complex cooling and remixing process than when they are in a critical mixture \cite{Siggia1979}, and this prevents the engine from working: the engine is obstructed by a bubble rich in one of the two phases created and stabilized around the particle.

\begin{figure*}
\includegraphics[width=0.7\textwidth]{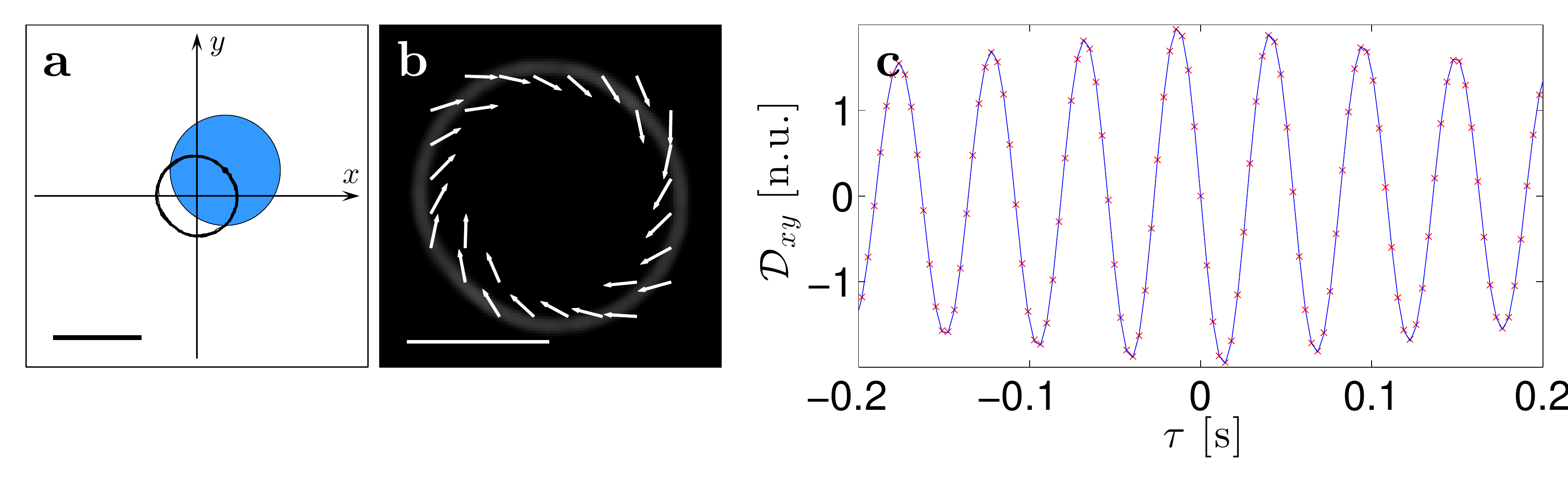}
\caption{
\textbf{Simulation of a critical engine.}
(a) Particle (blue circle) with corresponding $0.6\,{\rm s}$ trajectory (black line). The black bar corresponds to $1\,{\rm \mu m}$.
(b) Drift field (white arrows) and particle position distribution (background color, brighter colors represent higher probability distribution). The white bar corresponds to $1\,{\rm \mu m}$.
(c) Differential cross-correlation function $\mathcal{D}_{xy}$ in the $xy$-plane. 
}
\label{fig4}
\end{figure*}

\subsection{Numerical simulations}

In order to gain a deeper understanding of the mechanism responsible for the behavior we observed experimentally, we also investigate the dynamics of the system numerically (see also ``Numerical simulations" in the methods). We consider an absorbing hydrophilic particle held by an optical trap. We assume that the particle is slightly asymmetric in its shape and material properties (non-uniform absorption and surface roughness). The optical forces acting on the particle are calculated using generalized Mie theory \cite{jones2015optical}. The temperature increase in the surroundings of the particle is calculated by determining the absorption of the laser light by the iron oxide inclusions \cite{Peterman2003} and by using the stationary heat equation to estimate the ensuing heat conduction \cite{Rings2010}. This temperature increase causes a local demixing of the binary solution near the particle resulting in an increase in the local concentration of water, because of the hydrophilicity of the particle's surface. The diffusiophoretic drift and torque acting on the particle are then calculated from the slip velocity field generated around the particle \cite{Wurger2015}. Figure~\ref{fig4}a shows a typical trajectory obtained from simulations, which features the rotational motion observed also in experiments (compare Figs.~\ref{fig2}j and \ref{fig3}g); the corresponding drift field (Fig.~\ref{fig4}b) and $\mathcal{D}_{xy}$ (Fig.~\ref{fig4}c) are also in good agreement with the experiments (compare Figs.~\ref{fig2}k and \ref{fig3}h, and Figs.~\ref{fig2}l and \ref{fig3}i, respectively).

\subsection{Critical engine performance}

\begin{figure*}
\includegraphics[width=0.7\textwidth]{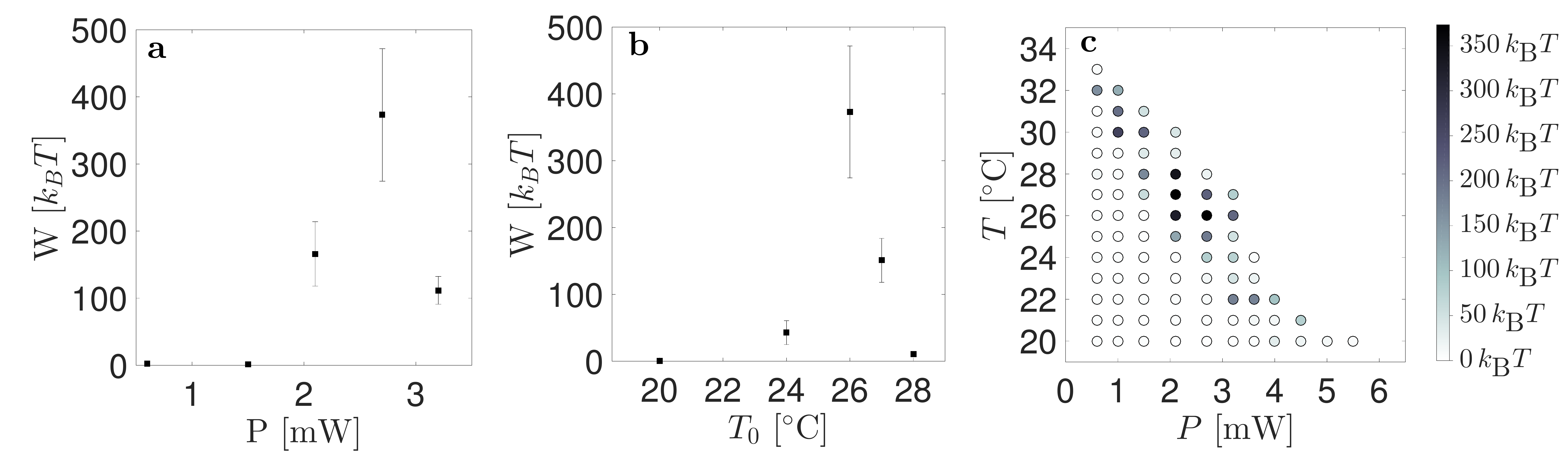}
\caption{
\textbf{Work dependence on laser power and ambient temperature.}
Work $W$ as a function of (a) laser power $P$ ($T_{0} = 26^{\circ}{\rm C}$) and (b) ambient temperature $T_{0}$ ($P = 2.7\,{\rm mW}$). (c) Work as a function of $P$ and $T_0$. The circles represent the performed measurements. $W$ features a maximum for $P=2.7\,{\rm mW}$ and $T_{0}=26^{\circ}{\rm C}$.
}
\label{fig5}
\end{figure*}

The rotational motion of the particle around the beam waist is intrinsically non-conservative and corresponds to a rotational force field. Through the analysis of $\mathcal{D}_{xy}$ (see ``Data analysis" in the methods), it is possible to evaluate the work performed by the particle during a single rotation.
The work performed during a single rotation about the $z$-axis can be expressed as \cite{Pesce2011}
\begin{equation}\label{eqn:work}
W = {2\pi}\, \frac {\Omega_{xy}}{\omega_\rho}\,k_{\rm B} T,
\end{equation}
where the work carried out by the rotating particle is directly proportional to the rotation frequency $\Omega_{xy}$ associated with the non-conservative force field and inversely proportional to the transverse relaxation frequency $\omega_\rho$ of the optical trap. The efficiency of the engine is given by $\eta_{\rm eff}=\frac{P}{\Omega_{xy}W}$.
The amount of work performed by the particle is plotted as a function of $P$ and $T_{0}$ in Fig.~\ref{fig5}. The maximum of $W = 373\,k_{\rm B} T$ and $\eta_{\rm eff}=1.16\cdot10^{-14}$ are reached for $P=2.7\,{\rm mW}$ and $T_{0}=26^{\circ}{\rm C}$. 
The amount of work performed by this critical engine exceeds that of colloidal heat engines such as the Brownian Carnot engine ($W_{\rm max}=5\,k_{\rm B}T$) \cite{Martinez2016} and the micro-metre sized heat engine ($W_{\rm max}=0.3\,k_{\rm B}T$) \cite{Blickle2012}. Its efficiency is comparable to the efficiency of a rotating object driven by the transfer of the angular momentum from a circularly polarized beam ($\eta_{\rm{eff}}\sim10^{-14}$) \cite{Friese1998} or by thermophoresis on asymmetric gears ($\eta_{\rm{eff}}\sim10^{-13}$) \cite{Maggi2015}.
We remark that, even though in our case the work done by the particle is immediately dissipated as heat into the fluid, it remains in principle accessible, e.g., by attaching a load to the particle. 

\section{Discussion}

We have realized a micron-sized critical engine that can extract work from the criticality of a system. In our realization the efficiency of the critical engine can be tuned by adjusting the incident laser power on the particle, the temperature of the ambient, or the criticality of the binary mixture. Compared to other micron-sized engines developed in the last years \cite{QuintoSu2014, Blickle2012, Martinez2016}, this critical engine has the advantages of not relying on the transfer of external angular momentum and of working at room temperature in contact with a single heat reservoir. The work performed per cycle by this engine exceeds those of other microscopic heat engines by orders of magnitude, while its efficiency is comparable to micron-sized engines driven by external angular momentum or thermophoresis \cite{Blickle2012,Martinez2016,Maggi2015}. 

Importantly, since many natural and artificial systems are tuned near criticality, the working principle of the microscopic engine that we describe here can be exploited in a diverse set of applications and be used to explain how natural phenomena work (e.g. molecular motors acting within a cellular membrane). Considering the wide range of systems that can be tuned near criticality, we can exploit any other order parameter tuned near criticality in a given system, such as pH-value and particle concentration \cite{Tagliazucchi2015}.
Since phase separations have been already widely found inside the human body \cite{Digel2006,Liu2004} and some of them are known to be sources of diseases such as protein condensations \cite{jacobs2017phase}, new biocompatible engines can be designed based on our critical engine that could be able to perform medical surgeries non-invasively such as the treatment of arteriosclerosis.

\section{Methods}

\subsection{Setup}

The schematic of the setup is shown in supplementary Fig.~S1. The optical tweezers is built on a homemade inverted microscope. The trapping laser beam ($\lambda = 976\,{\rm nm}$) is focused through an oil-immersion objective ($100\times$, ${\rm NA}=1.30$).

\setcounter{figure}{0}
\makeatletter 
\renewcommand{\thefigure}{S\@arabic\c@figure}
\makeatother

\begin{figure}
\includegraphics[width=0.3\textwidth]{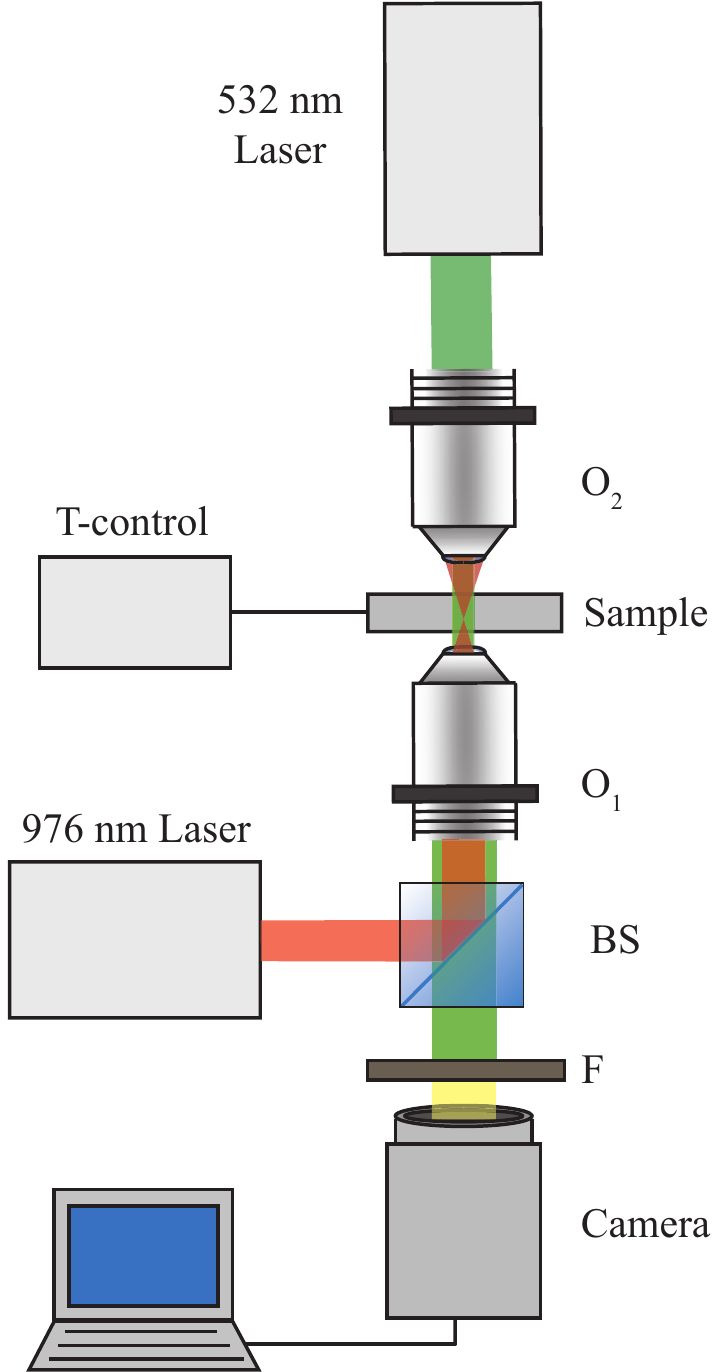}
\caption{
\textbf{Schematic of the experimental setup.}
The trapping laser ($\lambda= 976\,{\rm nm}$) is reflected by the dichroic beam splitter (BS) onto objective O$_{1}$ ($100\times$, ${\rm NA}=1.30$), which focuses the light inside the temperature-stabilized sample. The whole sample is illuminated by a green laser focused by objective O$_{2}$ ($20\times$, ${\rm NA}=0.45$); the forward-scattered light is filtered  (filter F) to eliminate the excitation laser light and projected onto a camera.
}
\label{sfig1}
\end{figure}

The sample chamber is prepared using a microscopic slide with a cavity (liquid volume $V=30\,{\rm \mu l}$) and sealed using a coverslip. The whole sample is temperature-stabilized using a copper-plate heat exchanger coupled to a water circulating bath (T100, Grant Instruments) with $\pm 50\,{\rm mK}$ temperature stability. Two Peltier elements (TEC3-6, Thorlabs) placed on the trapping objective permit us to fine-tune ($\pm 3\,{\rm mK}$) the temperature using a feedback controller (TED4015, Thorlabs).

A second laser ($\lambda=532\,{\rm nm}$) is focused on the sample by a $20\times$ objective (${\rm NA}=0.45$) and excites the fluorescent dye Rhodamine B (C.I. 45170, Merck) dissolved in the solution. The emission peak of this dye is at $\lambda \approx 600\,{\rm nm}$. A CMOS camera (DCC1645C, Thorlabs) records the light emitted by the sample at a frame rate of $296\,{\rm fps}$.

The recorded videos are analyzed using standard digital video microscopy algorithms to obtain the three-dimensional position of the particle. We have found that the particle predominantly moves in the $xy$-plane. We have therefore neglected the information about the $z$-position in our analysis.

\begin{figure*}
\includegraphics[width=.8\textwidth]{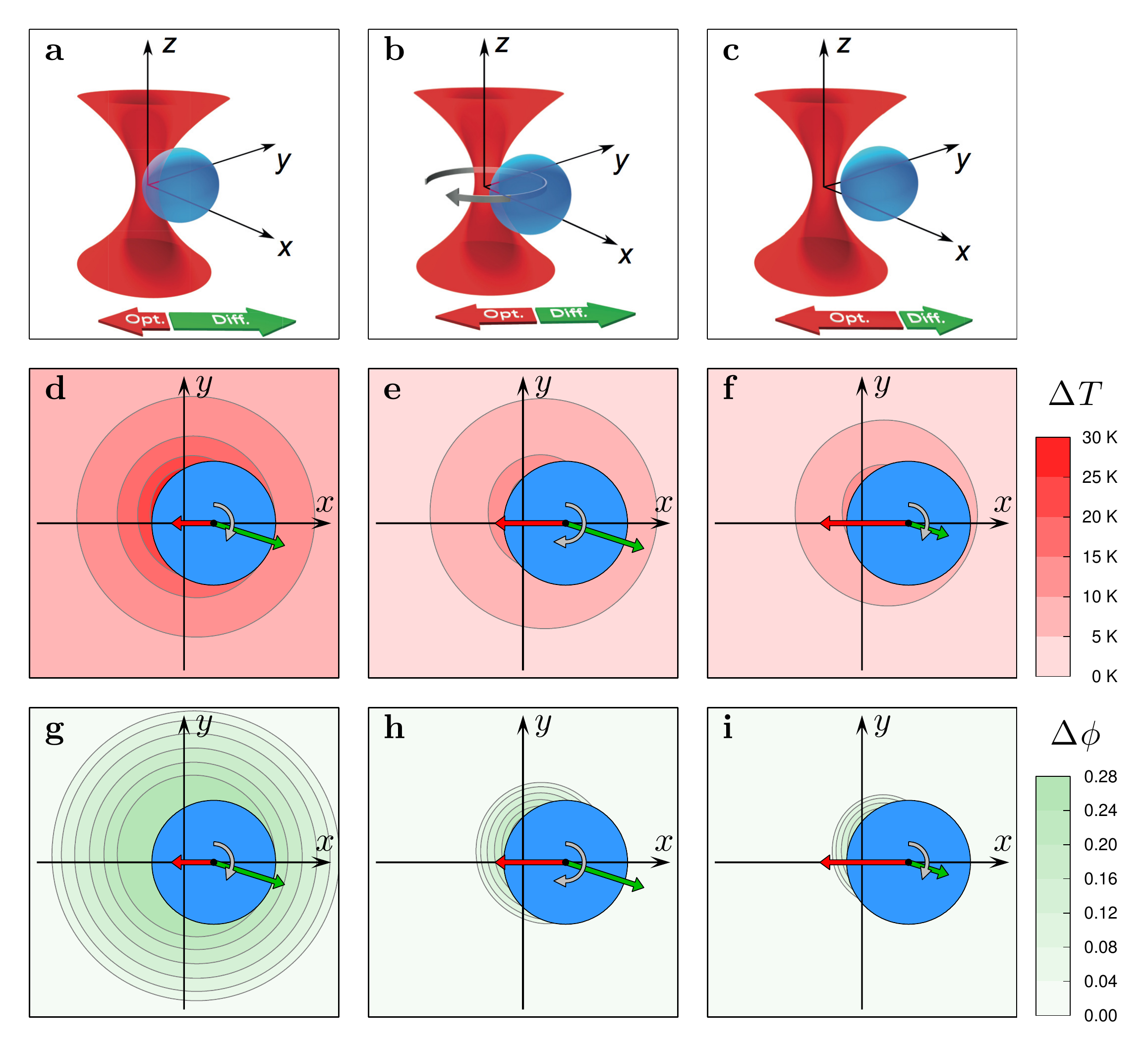}
\caption{
\textbf{Simulation of a critical engine.}
(a-c) Absorbing particle in a critical mixture held at various radial distances from the center of an optical trap, and corresponding (d-f) temperature profiles (the isotemperature lines  are spaced by $5\,{\rm K}$) and (g-i) concentrations profiles (the isoconcentration lines are spaced by 0.04) in the $xy$-plane ($z=0$). 
As the distance from the center of the trap increases, the drifts due to the optical forces (red arrows) increase and the diffusiophoretic drifts (green arrows) decrease. 
The presence of small asymmetries in the temperature and demixing profile around the particle make it rotate around the optical axis; the corresponding tangential force (gray arrow) is maximized at the equilibrium configuration where the optical-force-induced drift balances the diffusiophoretic drift.
(a), (b), and (c) correspond to Fig.~1(a), 1(c), and 1(b) respectively.
}
\label{sfig2}
\end{figure*}

\subsection{Data analysis}

Each trajectory is analyzed by calculating the corresponding drift field, differential cross-correlation function $\mathcal{D}_{xy}$, and work.

The velocity drift fields are derived from the measured trajectory as
\begin{equation}\label{eqn:driftvel}
{\bf v}({\bf r})=\frac{1}{\Delta t}\left<{\bf r}_{n+1}-{\bf r}_{n} | {\bf r}_{n} \approx {\bf r} \right>,
\end{equation}
where ${\bf r}_{n}$ is the position of the particle in the $xy$-plane and $\Delta t$ is the time interval between subsequent positions of the particle. The calculated magnitude and direction of the local velocity is indicated by white arrows in Figs.~\ref{fig2} and \ref{fig3}.

The differential cross-correlation function $\mathcal{D}(\tau)$ of the particle motion is calculated as  \cite{Volpe2006,Volpe2007}:
\begin{equation}\label{dccf}
\mathcal{D}_{xy}(\tau)=C_{xy}(\tau) - C_{yx}(\tau),
\end{equation}
where $C_{xy}(\tau)=\frac{\langle x(t)\ y(t+\tau)\rangle}{\sqrt{ \langle x(t)^2\rangle \langle y(t)^2\rangle }}$ and $C_{yx}(\tau)=\frac{\langle y(t)\ x(t+\tau)\rangle}{\sqrt{ \langle x(t)^2\rangle \langle y(t)^2\rangle }}$. The theoretical expression of $\mathcal{D}_{xy}(\tau)$ for a trapped spherical particle rotating in a plane can be obtained from the Langevin equation in a non-homogeneous force field \cite{Volpe2007} as
\begin{equation}\label{dcfg}
\mathcal{D}_{xy}(\tau)=2D \ \frac {e^{-\omega_\rho |\tau|}}{\omega_\rho} \sin \left( \Omega_{xy} \tau \right),
\end{equation}
where $D$ is the diffusion coefficient, $\omega_\rho$ is the radial relaxation frequency of the particle in the optical trap (obtained from the autocorrelation function \cite{jones2015optical}), and $\Omega_{xy}$ is the rotational frequency of the particle in the $xy$-plane.
The value of $\Omega_{xy}$ is obtained by fitting the experimental $\mathcal{D}_{xy}(\tau)$ (Eq.~(\ref{dccf})) to the theoretical expression in Eq.~(\ref{dcfg}). The resulting  hydrodynamic viscous torque is then:  
\begin{equation} 
\overline{\mathbf{\Gamma}}_{xy}= \overline{\mathbf{r}\times \mathbf{F}_{\rm drag}} \ = \gamma\, \overline{\mathbf{r}\times \mathbf{r}\times \Omega_{xy}}=\gamma \ \Omega_{xy}\ \sigma^2_{xy}\ \mathbf{\hat z},
\end{equation} 
 
\noindent where $\mathbf{r}$ is the position of the particle, $\gamma=6\pi\eta R$ is the friction coefficient, defined by Stokes law and related to the medium viscosity $\eta$ and to the particle's radius $R$, and $\sigma^2_{xy}$ is the variance of the particle's position in the plane orthogonal to the torque. 

The work performed during a single rotation about the $z$-axis can be expressed as \cite{Pesce2011}: 
\begin{equation}\label{eqn:work}
W= \int_{0}^{2\pi}\, \overline{\Gamma}_{xy}\, d\theta={2\pi}\, \frac {\Omega_{xy}}{\omega_\rho}\,k_{\rm B}T.
\end{equation}

\subsection{Numerical simulations}

The motion of the particle is simulated using a standard finite-difference algorithm based on a three-dimensional Langevin equation describing the motion of the particle under the action of Brownian motion, optical forces, and diffusiophoretic drifts and torques \cite{jones2015optical}.
The particle is modeled as a silica microsphere (radius $R=1.25\ \mu{\rm m}$) with iron-oxide inclusions (25\% of total weight) distributed inhomogeneously near the surface of the particle.
The optical force on the particle is calculated using generalized Mie theory (particle refractive index $n_{\rm p}=1.46$, medium refractive index $n_{\rm m}=1.38$ \cite{Grattoni1993}) and assuming a linearly polarized Gaussian beam (wavelength $\lambda_{\rm 0}=976\ {\rm nm}$) focused through a high-NA objective (${\rm NA}=1.30$) \cite{jones2015optical}.
The temperature increase as a function of position, $\Delta T(\mathbf{r})$, is obtained using Fourier's law of heat conduction \cite{Peterman2003}:
\begin{equation}
\nabla^{2}(\Delta T(\mathbf{r}))=-\frac{\alpha}{C} I(\mathbf{r}),
\end{equation}
where $I(\mathbf{r})$ is the light intensity, $\alpha$ is the absorption coefficient of iron (for iron $\alpha=2.43\cdot10^7\,{\rm m^{-1}}$, we assume no absorption in silica and water--2,6-lutidine), and $C$ is the thermal conductivity (for iron $C=73\,{\rm Wm^{-1} K^{-1}}$, for silica $C=1.4\,{\rm Wm^{-1} K^{-1}}$, for water--2,6-lutidine $C=0.39\,{\rm Wm^{-1} K^{-1}}$).
Where $T_0+\Delta T(\mathbf{r})>T_{\rm c}$, a concentration gradient $\Delta \phi (\mathbf{r})$ is induced, which, in proximity of the critical temperature  $T_{\rm c}$, is \cite{Wurger2015}
\begin{equation}
\Delta \phi(\mathbf{r})= \sqrt{\frac{T_0+\Delta T(\mathbf{r})-T_{\rm{c}}}{K}},
\end{equation}
where $K$ is a constant. The concentration gradient $\Delta \phi (\mathbf{r})$ generates a slip velocity field ${\mathbf v}_{\rm s} (\mathbf{r})$ in the layer around the particle. The diffusiophoretic drift is then ${\mathbf v}_{\rm p} = -\langle {\mathbf v}_{\rm s} (\mathbf{r}) \rangle$ \cite{Wurger2015}, while the diffusiophoretic torque is ${\mathbf T}_{\rm ph} =   \langle {\mathbf r}_{\rm s} \times (- \gamma {\mathbf v}_{\rm s}) \rangle$, where ${\mathbf r}_{\rm s}$ is the vector connecting the center of mass of the particle to the particle's surface.

\section{Acknowledgments}

We thank Andrea Gambassi and Klaus Kroy for fruitful discussions about the theory, and Christian Schmidt from Microparticles GmbH for insightful discussions about the particles.
This work was partially supported by the ERC Starting Grant ComplexSwimmers (grant number 677511) by Vetenskapsrådet (grant number 2016-03523]. AC acknowledges partial support from T\"ubitak (grant number 115F401). 
 
\section{Author contributions}

FS implemented the experimental setup, performed the measurements, analysed the experimental data, and contributed to the simulations.
AM contributed to the implementation of the experimental setup, the measurements and the analysis of the experimental data.
AC performed the simulations.
LB provided guidance for the simulations and interpretation of the data.
FC contributed to the interpretation of the experimental data and their simulation.
GV conceived and supervised the work.
All authors participated in discussing the results and in writing the article.

\bibliographystyle{unsrt}
\bibliography{biblio}

\end{document}